\documentclass[12pt]{iopart}

\usepackage{iopams}
\usepackage{graphicx}
\usepackage{appendix}
\usepackage{color}

\begin{document}

\title{Metal enhanced fluorescence in rare earth doped plasmonic core-shell nanoparticles}

\author{S. Derom$^{1}$, A. Berthelot$^{2}$, A. Pillonnet$^{2}$, O. Benamara$^{2}$, A.M. Jurdyc$^{2}$, C. Girard$^{3}$ and G. {Colas des Francs}$^{1}$}

\address{
$^{1}$Laboratoire Interdisciplinaire Carnot de Bourgogne (ICB), UMR 6303 CNRS-Universit\'e de Bourgogne, 
9 Av. A. Savary, BP 47 870, F-21078 Dijon, France \\
$^{2}$Institut Lumi\`ere Mati\`ere, UMR 5306 Universit\'e de Lyon 1-CNRS, Universit\'e Lyon, Villeurbanne F-69622, France \\
$^{3}$Centre d'Elaboration de Mat\'eriaux et d'Etudes Structurales (CEMES), 
CNRS, 29 rue J. Marvig, BP 94347, F-31055 Toulouse Cedex 4, France
}

\ead{gerard.colas-des-francs@u-bourgogne.fr}

\submitto{\NT}
%\maketitle

\begin{abstract}
We theoretically and numerically investigate metal enhanced fluorescence of plasmonic core-shell nanoparticles doped with 
rare earth (RE) ions. Particle shape and size are engineered to maximize the average enhancement factor (AEF) of the overall doped shell. 
We show that the highest enhancement (11 in the visible and 7 in the near-infrared) 
are achieved by tuning either the dipolar or quadrupolar particle resonance to the rare earth ions excitation wavelength. Additionally, the calculated AEFs are compared to experimental data reported in the literature, obtained in similar conditions 
(plasmon mediated enhancement) or when a metal-RE energy transfer mechanism is involved.  
\end{abstract}

\maketitle

\section{Introduction}
Rare earth (RE) ions are widely studied for numerous optical applications such as solar cells \cite{Timmerman:2008}, optical amplification, biolabelling \cite{Bouzigues:2011} but also photodynamic therapy of cancer \cite{Wang:2011}. Although they present high quantum efficiencies, they could suffer from low absorption 
cross-section \cite{Bunzli:2007} (around $10^{-20}cm^{2}$) so that metal enhanced fluorescence (MEF) has been proposed to improve their emission properties. 

Metal enhanced spectroscopies rely on the excitation and/or emission exaltation by coupling emitters to a plasmonic particle. 
This has been extensively studied, notably for Tip-- or Surface-- Enhanced Raman Scattering (TERS/SERS) \cite{Pettinger:2010} 
or dye fluorescence enhancement \cite{Fort-Gresillon:2008,Giannini-FrenadezDominguez-Heck-Maier:2011}.  Practically, the highest signal enhancement is achieved for low initial absorption cross--section {\it and} quantum yield, 
since both excitation and emissions processes are enhanced.  Sun {\it et al}  described SERS as photoluminescence enhancement 
in the limit of null initial absorption cross section and quantum yield \cite{Sun-Khurgin-Tsai:2012}. 
Therefore, SERS presents the highest enhancement (up to $10^6$) \cite{Fang-Seong-Dlott:2008} whereas MEF is typically of few tens only \cite{bharadwaj07b}. 
{Nevertheless, the results achieved for dye molecules cannot be directly transposed to rare-earth ions which present extremelly low absorption cross-section but quantum efficiency close to unity on the contrary to dyes which present high absorption cross-section, and generally lower quantum efficiency. In addition, it is worthwhile to note that lanthanides luminescence is also extremelly sensitive to the surroundings so that identifying the role of plasmon is a difficult task.}

A large number of works,  mainly experimental \cite{Strohhofer-Polman:2002,Mertens-Polman:2006,Marques-Almeida:2007,Aslan-Wu-Lakowics-Geddes:2007,Ma-Dosev-Kennedy:2009,Kassab-Araujo:2010,Som-Karmakar:2011,vanWijngaarden-Meijerink:2011,Reisfeld-Gaft:2011,Deng-Goldys:2012,Rivera-Nunes-Marega Jr:2012,Amjad:2013}, but also theoretical \cite{Malta-CoutoDosSantos:1990,Greffet:2009,Fischer-Goldschlmidt:2012}, 
have been realized in order to probe the possibility to enhance optical properties of rare earth ions placed near metal 
nanoparticles. Some recent works have highlighted that gold or silver metal nanoparticle could change the 
selection rules of rare earth ions emission \cite{Kassab-Araujo:2010,Karaveli:2011} 
but the localized plasmon contribution remains in discussion. 
Indeed, it is very difficult to separate respective roles of plasmon 
and energy transfer in the observed enhancement of RE ions luminescence \cite{Eichelbaum:2009}. 
Plasmon mediated enhancement relies on antenna effect that increases the excitation field and/or radiative emission rate 
\cite{Busson-Bidault:2012}, whereas energy transfer is a dipole-dipole F\"orster-like mechanism between the metal 
nanocrystal (donor or sensitizer) 
and RE ions (acceptor) \cite{Strohhofer-Polman:2002}. It has been observed that energy transfer is at the origin of luminescence enhancement 
near small gold and silver clusters composed by few atoms (crystal size of few nm) \cite{Eichelbaum:2009,Maurizio:2011}. 
However, larger metal particles (typically few tens of nm) are generally preferred for plasmon-enhanced 
fluorescence\cite{Aslan-Wu-Lakowics-Geddes:2007,vanWijngaarden-Meijerink:2011,Reisfeld-Gaft:2011,Deng-Goldys:2012,Amjad:2013}. 
{
Recently, Ma {\it et al} measured 3-fold and 24-fold enhancement factor for $\mathrm{Eu^{3+}}$ doped silver core-shell nanoparticles with 
20 nm core/15 nm shell and 9 nm core/11 nm shell, respectively \cite{Ma-Dosev-Kennedy:2009}. However, they used 
$\lambda_{exc}=260$ nm excitation wavelength far from the dipolar resonance 
so that we think that the enhancement mechanism is different from plasmon enhanced spectroscopy and most likely originates 
from energy transfer. Two others groups reported 3 fold enhancement for $\mathrm{Eu^{3+}}$ ions coupled to about 30 nm silver particles 
when excited close to the silver dipolar resonance \cite{Reisfeld-Gaft:2011,Amjad:2013}.
In a former work, Malta and Couto dos Santos have made a rough estimation 
of the possible emission enhancement for Europium ions doped glasses containing silver nanoparticles  
\cite{Malta-CoutoDosSantos:1990}. They estimated up to 50-fold maximum {\it local} enhancement factor for 30 nm silver particles 
but not conclude about the global enhancement of the overall doped shell.}

The purpose of this work is to theoretically determine and optimize the plasmon contribution to the luminescence enhancement 
in plasmonic nanoparticle doped with lanthanides. 
Particular attention is devoted to describe an ensemble of  rare earth  ions coupled to a metallic nanoparticle 
instead of single emitter coupled to one particle as generally done since we are interested in the optical response of the whole doped 
nanostructure. In a first time, we investigate the role of the localized plasmons supported by the nanoparticle and determine 
the optimal particle resonance position compared to the emitter absorption and emission peak. 
To this end, we define in section \ref{sect:AEF}, the average fluorescence factor 
for a doped plasmonic core-shell particle of arbitrary shape. 
For comparison purpose, we illustrate the enhancement mechanism by considering a laser dye, 
namely the Rhodamine 6G (Rh6G) coupled to a spherical metal particle \cite{Xiong:2012}. 
In a second time, in section \ref{sect:RareEarth}, we investigate 
the fluorescence enhancement for rare earth ions emitting in the visible 
or the near infrared and placed in the shell of core-shell particles with metal core. 
We will thus estimate the achievable plasmonic enhancement.

\section{Surface enhanced fluorescence} 
\label{sect:AEF}
In this section, we derive a general expression for the average fluorescent enhancement factor (AEF) near a metal particle. 
To this aim, we extend the work of Liaw {\it et al} \cite{Liaw-Kuo:2010} to an arbitrary geometry.
We first derive a closed form expression of the enhancement factor $\alpha({\bf r})$ for randomly oriented emitters 
near a metal particle. 
AEF is then achieved by numerically averaging $\alpha({\bf r})$ over the doped shell volume. 
\subsection{Enhancement factor of randomly oriented emitters}
\begin{figure}[h!]
\begin{center}
\includegraphics[width=16cm]{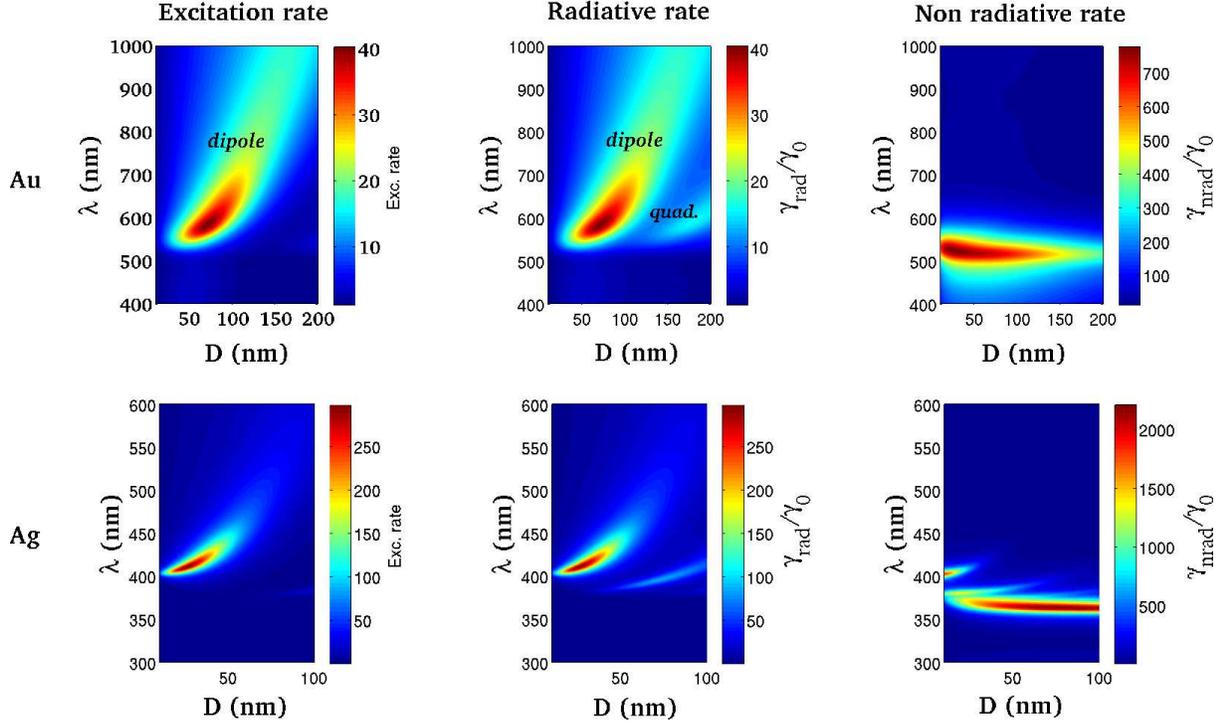}
\caption{\label{fig:AuAgFluoDisp}Excitation (left), radiative (middle) and non-radiative (right) rates 
near a metal spherical particle as a function of wavelength and particle diameter $D$. 
Top (bottom) line refers to gold (silver) particle. 
The dipolar emitter is located 5 nm from the particle surface and perpendicular to it. 
The optical index of the embedding matrix is n=1.5. 
Metal dielectric constant are taken from Johnson and Christy \cite{Johnson-Christy:1972}.}  
\end{center}
\end{figure}

Let us first consider a single fluorescent system with a transition dipolar moment of arbitrary orientation 
${\bf p}=p_0(\sin \alpha \cos \beta,\sin \alpha \sin \beta, \cos \alpha)$ and intrinsic quantum yield $\eta_0$. 
If the incident field is ${\bf E_0}$, the fluorescent signal in absence of plasmonic particle is 
$p_0^2\left \vert{\bf E}_0\right \vert ^2 \eta_0$. The plasmonic particle modifies: 
\begin{itemize}
 \item the excitation rate $\pi({\bf r},\omega_{exc})={\left \vert{\bf p} \cdot {\bf E}({\bf r},\omega_{exc})\right \vert ^2}/{p_0^2\left \vert{\bf E}_0\right \vert ^2}$ where ${\bf E}({\bf r},\omega_{exc})$ 
refers to the excitation field at the emitter location ${\bf r}$
and excitation angular frequency $\omega_{exc}$
\item and the emitter quantum yield $\eta({\bf r},\omega_{em})=\gamma^{rad}/(\gamma^{rad}+\gamma^{NR})$ 
at the emission angular frequency $\omega_{em}$.
\end{itemize}
As an example, we present on Fig.~\ref{fig:AuAgFluoDisp} the excitation and decay rates calculated for an emitter close to 
gold or silver beads. The emitter is perpendicular to the particle surface since stronger effects are expected for this orientation. 
The excitation and radiative rates follow the dipolar plasmon mode dispersion 
for small particle diameters. Large particles support a leaky quadrupolar mode so that the radiative rate couples to this mode. 
Note that the excitation rate weakly follows the quadrupolar mode dispersion since the emitter is not located on a mode lobe. 
Finally, the non radiative rate originates from coupling to high order modes 
so that it presents a flat dispersion curve \cite{OptExpGCF:2008}. 
The low order modes are well-separated in case of a silver particle due to lower losses 
(compare the non radiative rate near gold and silver structures).  
Last, we observe that the dipolar plasmon resonance can be tuned from $\lambda=525~\mathrm{nm}$ (well-defined resonance) 
to $\lambda\approx 900~\mathrm{nm}$ (large resonance) 
for gold beads diameters varying from $D=10~\mathrm{nm}$ to $D=200~\mathrm{nm}$. For silver nanoparticles diameters between $10~\mathrm{nm}$ and $100~\mathrm{nm}$, 
the dipolar resonance wavelength varies from $\lambda=400~\mathrm{nm}$ to $\lambda=530~\mathrm{nm}$. 

Finally, both the excitation and emission rates depend on the emitter location and orientation in presence of plasmonic nanostructures. 
The enhancement factor expresses
\begin{equation} 
\alpha_{\bf p}({\bf r},\omega_{exc},\omega_{em}) = \frac{\left \vert{\bf p} \cdot {\bf E}({\bf r},\omega_{exc})\right \vert ^2  
\eta({\bf r},\omega_{em})}
{p_0^2\left \vert{\bf E}_0\right \vert ^2 \eta_0} \,.
\end{equation}

In case of arbitrary orientation, the quantum yield expresses \cite{JCPGCF:2002}
\begin{eqnarray}
\eta &=& \sin^2 \alpha \cos^2 \beta ~  \eta_x + \sin^2 \alpha \sin^2 \beta   ~\eta_y +\cos^2 \alpha ~ \eta_z \\
&&+\sin^2 \alpha \cos \beta \sin \beta ~\eta_{xy}+ \sin \alpha \cos \alpha \cos \beta  ~ \eta_{xz}
+\sin \alpha \cos \alpha \sin \beta ~\eta_{yz} \,,
\nonumber
\end{eqnarray}
where $\eta_i=\Gamma_{rad,i}/(\Gamma_{rad,i}+\Gamma_{NR,i})$ is the quantum yield associated to the $i$-orientation (i=x,y or z). 
$\eta_{ij}$ (i,j=x,y or z) refer to crossed-terms that have to be considered for oblique dipole moments 
(see Refs. \cite{JCPGCF:2002,PRELevequeGCF:2002} for details). 
Assuming randomly oriented emitters at location ${\bf r}$, we calculate the mean enhancement factor over all the possible orientations 
\begin{equation}
\alpha({\bf r},\omega_{exc},\omega_{em}) =\frac{3}{4 \pi} \int_{\alpha=0}^{\pi}  \int_{\beta=0}^{2\pi}  
\alpha_{\bf p}({\bf r},\omega_{exc},\omega_{em}) ~ \sin \alpha ~d \alpha ~d \beta \,,
\end{equation}
where the factor $3$ ensures a unit enhancement factor for isolated emitters excited with a linearly polarized field. 
Interestingly enough, the integration over the dipole orientation is analytical. After a few algebra, we achieve
\begin{eqnarray}
\label{eq:alpha}
&&\alpha({\bf r},\omega_{exc},\omega_{em}) \\
&&=\frac{\eta_x(3E_x^2+E_y^2+E_z^2)+\eta_y(E_x^2+3E_y^2+E_z^2)+\eta_z(E_x^2+E_y^2+3E_z^2)}{5 \left\vert{\bf E}_0\right \vert ^2 \eta_0} \,.
\nonumber
\end{eqnarray}

It is also useful to derive this expression in spherical coordinates ${\bf r}=(r,\theta,\phi)$. 
With subscripts $//$ and $\perp$ indicating an orientation parallel or perpendicular to the particle surface, it writes
\begin{equation}
\label{eq:alphaMie}
\alpha({\bf r},\omega_{exc},\omega_{em})= \frac{\eta_{//}(2E_r^2+4E_{\theta}^2+4E_{\phi}^2)+\eta_{\perp}(3E_r^2+E_{\theta}^2
+E_{\phi}^2)}{5\left\vert{\bf E}_0\right \vert ^2 \eta_0}  \,,
\end{equation}
that leads to an analytical expression for (homogeneous, core--shell or onion--like) spherical particles, 
thanks to Mie expansion \cite{Kim-Leung-George:1988,Sinzig-Quinten:1995}. 
In the following, $\alpha({\bf r},\omega_{exc},\omega_{em})$ is referred as the local fluorescence enhancement. 

\begin{figure}[h!]
\begin{center}
\includegraphics[width=14cm]{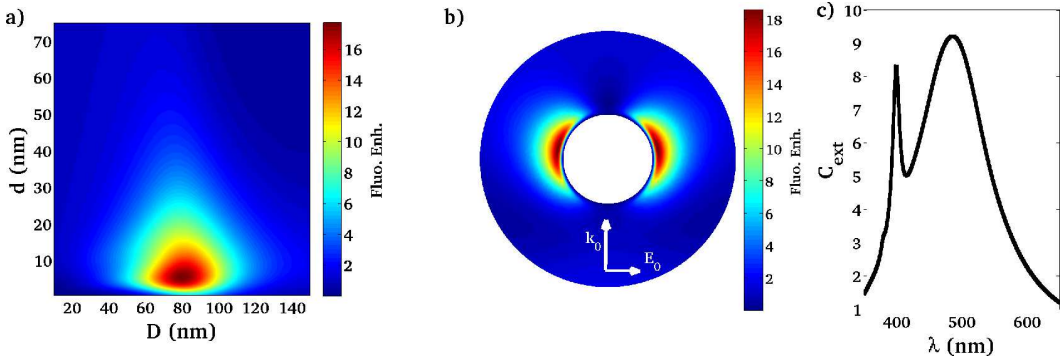}
\caption{\label{fig:AgRh6G}Fluorescence enhancement near a silver particle for randomly oriented molecules. 
a) As a function of the particle diameter and dye-particle distance (the molecules are along the incident field polarization axis). 
b) Near a 80 nm silver particle. The excitation and emission wavelength are 
$\lambda_{exc}=532 \mathrm{nm}$ and $\lambda_{em}=560 \mathrm{nm}$, respectively. The embedding medium is PMMA (optical index $n=1.5$).
c) Extinction efficiency of a 80 nm silver particle.}
\end{center}
\end{figure}  
Obviously, the choice of the particle size, shape and material depends on the fluorescent system. 
In case of dye molecules, with a low Stokes shift between the absorption and emission wavelength, the most efficient 
fluorescent enhancement is achieved when the particle dipolar resonance overlaps both the excitation and emission wavelengths \cite{bharadwaj07b,Reineck:2013}. 
For instance, let us consider Rhodamine 6G. We will compare rare-earth doped nanoparticle to this reference system. 
The absorption and emission peaks are near 
$\lambda_{exc}=532~\mathrm{nm}$ and $\lambda_{em}=560~\mathrm{nm}$, respectively. Figure \ref{fig:AgRh6G} 
shows the local fluorescent enhancement for randomly oriented Rh6G 
molecules dispersed in polymethyl metacrylate (PMMA) near silver spherical particles. The maximum fluorescence enhancement occurs for 
dye molecules coupled to a 80 nm particle (Fig.~\ref{fig:AgRh6G}a). Such large silver bead support a dipolar plasmon about 
$\lambda\approx 500 \mathrm{nm}$ with a broad resonance width (and a quadrupolar mode at $\lambda=400~\mathrm{nm}$, see Fig.\ref{fig:AgRh6G}c). 
Therefore both excitation field and radiative rates are significantly enhanced by coupling 
to the dipolar mode. Meanwhile, the non-radiative rate remains limited due to bad coupling to high order modes 
that appear at lower wavelength (see also Fig.~\ref{fig:AuAgFluoDisp}). 
Figure \ref{fig:AgRh6G}b) presents the local enhancement factor calculated 
near a 80 nm silver sphere. 
It follows the particle dipolar mode profile, with a maximum enhancement $\alpha \approx 18$ slightly shifted 
from the incident polarisation axis \cite{Liaw-Kuo:2010,Hartling-Reichenbach-Eng:2007}.

\subsection{Layer-- and shell--average enhancement factor}

Finally, the average enhancement factor of the whole doped volume $V_0$ can be numerically computed as 
\begin{eqnarray}
\label{eq:AEF}
AEF(\omega_{exc},\omega_{em})&=&\frac{1}{V_0} 
\int \! \! \! \! \! \int \! \! \! \! \! \int_{{\bf r} \in V_0} \alpha({\bf r},\omega_{exc},\omega_{em}) dV
\end{eqnarray}
In case of spherical particles, it is also useful to define an average enhancement factor associated to a doped shell layer 
\begin{eqnarray}
\label{eq:LayerAEF}
{\alpha}_{layer}(\omega_{exc},\omega_{em},{\bf r})&=&\frac{1}{4\pi} \int_{\theta=0}^{\pi}  \int_{\phi=0}^{2\pi}
\alpha({\bf r},\omega_{exc},\omega_{em})
~ \sin \theta ~d \theta ~d \phi
\end{eqnarray}
Note that average and layer fluorescence enhancement factors are normalized with respect to the doped volume so that they do not depend on 
the RE concentration but rather characterize the SPP mediated fluorescence enhancement.

\begin{figure}[h!]
\begin{center}
\includegraphics[width=6cm]{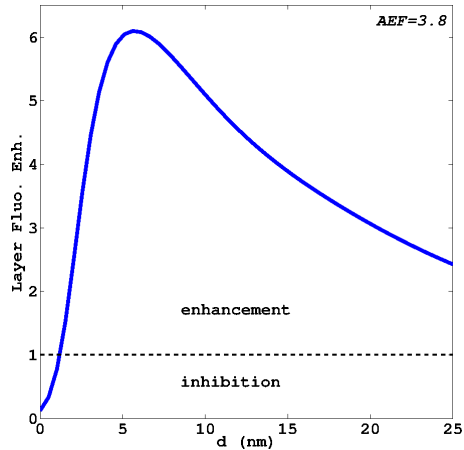}
\caption{\label{fig:AEFAgRh6G} Layer fluorescence enhancement of a Rh6G doped core-shell $\mathrm{Ag@SiO_2}$ (silver core 80 nm diameter). 
The horizontal line indicates the enhancement/inhibition threshold.
The average fluorescence enhancement of the whole doped layer (25 nm) is $AEF=3.8$.}
\end{center}
\end{figure}

Figure \ref{fig:AEFAgRh6G} shows the layer fluorescence enhancement ${\alpha}_{layer}$ and the average fluorescence factor 
of the dye doped core-shell particle. The overall $AEF \approx 4$ is optimized given the absorption and emission wavelengths 
of the dye molecule. This factor takes into account inhomogeneous excitation in the dipolar modes as well as the distance 
dependence of the emission rate. 
 
\section{Rare earth doped plasmonic core shell} 
 \label{sect:RareEarth}
  
In the previous section, we have introduced two important parameters in order to quantify 
plasmon/emitters interactions in a core-shell particle: layer-averaged enhancement factor $\alpha_{layer}$, and 
shell-averaged enhancement factor AEF. We have estimated maximum $AEF\approx 4$ achievable for a 
Rh6G doped core-shell $\mathrm{Ag@SiO_2}$ reference system. 
These factors will now allow us to characterize the effect of a metal nanoparticle on the luminescence of RE ions.
 
In this part, we study the luminescence enhancement for two rare earth ions used in different application domains: 
$\mathrm{Eu^{3+}}$ used for example as biolabel for its emission in the visible spectrum and $\mathrm{Er^{3+}}$, 
as main active medium for amplification at the telecom wavelength $1.55~\mu m$. 
Compared to the dye case, the situation could be rather different when considering rare earth ions fluorescence. 
Quantum efficiency of these emitters is close to unity, so plasmon resonance cannot much increase this factor. 
Lanthanide absorption, based on 4f transitions, in contrast, is very  weak (around $10^{-20}cm^{2}$). 
Therefore, the strongest exaltation is expected when matching the lanthanide absorption wavelength with the dipolar plasmon resonance. 
Emission of lanthanide could be far from excitation wavelength and then probably not 
much influenced by plasmon phenomenon \cite{Pillonnet-Jurdyc:2012} contrary to previous section where the dye excitation and emission peaks 
were within the dipolar resonance.

\subsection{Emission in the visible}

\begin{figure}[h!]
\begin{center}
\includegraphics[width=13cm]{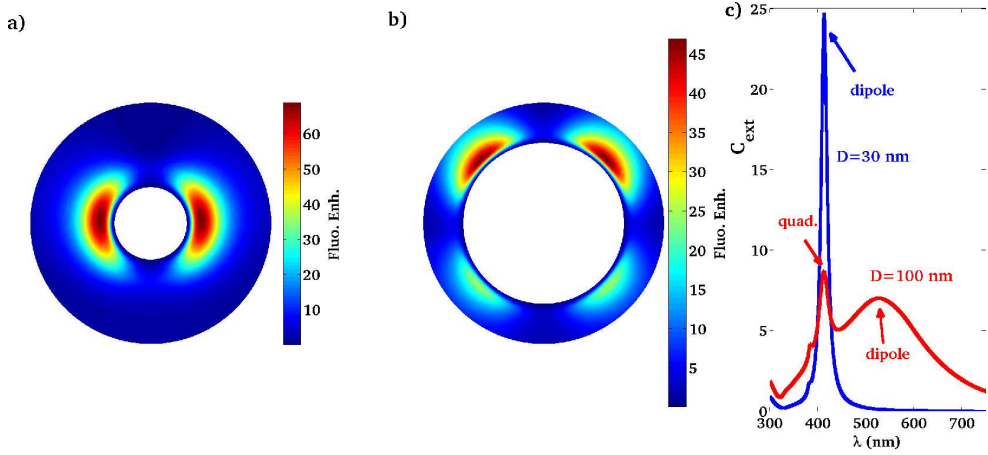}
\caption{\label{fig:AgEu}$\mathrm{Eu^{3+}}$ fluorescence enhancement near a 30 nm (a) or 100 nm (b) silver particle. 
c) Extinction efficiency of a 30 nm (blue line) or 
100 nm (red line) silver particle. The dipolar and quadrupolar resonance are indicated for the two particle diameters.
Excitation and emission wavelength are 
$\lambda_{exc}=415~ \mathrm{nm}$ and $\lambda_{em}=620~ \mathrm{nm}$, respectively. The embedding medium is $\mathrm{SiO_2}$ 
(optical index $n=1.5$).  }
\end{center}
\end{figure}

First we consider Europium system: this rare earth ion, usually UV/blue excited, 
has a maximum emission around 620 nm and a quantum yield close to 1. 
Since their excitation corresponding to 5D0-7F2 transition is in the blue part of the visible spectrum, 
we have chosen to use silver for the nanoparticle core. 

We investigate $\mathrm{Eu^{3+}}$/silver system in Fig.~\ref{fig:AgEu}, where rare earth ions ($\mathrm{Eu^{3+}}$) 
are included in a dielectric shell and coupled to a silver bead. 
We consider a silica matrix of same index as PMMA so that direct comparison with the previous dye molecule case is possible. 
This luminescent doped core-shell $\mathrm{Ag@SiO_2}$ particle can be chemically synthesized 
\cite{Aslan-Wu-Lakowics-Geddes:2007,vanWijngaarden-Meijerink:2011,Deng-Goldys:2012}. 
The excitation and emission wavelengths are $\lambda_{exc}=415~\mathrm{nm}$ and $\lambda_{em}=620~\mathrm{nm}$, respectively, 
corresponding to f-f intra-configurational transitions. As previously, we first determine the size of the  metal core that 
optimizes red luminescence of the europium ions. 
We find that the maximum local fluorescent enhancement ($\approx 70$) is obtained for a 30 nm silver core. We plot in Fig.~\ref{fig:AgEu}a 
the map of the fluorescence exaltation achieved near the 30 nm silver sphere. 
This enhancement is mainly an improvement of absorption process: we have a strong exaltation of the excitation field resonant 
with the dipolar resonance of metal nanoparticle. 
Since the emission wavelength is far from all the particle resonances (see the extinction efficiency in Fig.~\ref{fig:AgEu}c), 
decay rates are practically not affected by the presence of the plasmonic nanostructure. 
Table \ref{table:Rates} gathers the excitation and decay rates for dye-doped and RE-doped silver core-shell. 
For dye molecules, with small Stokes shifts, large particles leads to stronger effect since it corresponds to large resonances 
and coupling to the dipolar plasmon enhances both the excitation and radiative emission rates. 
Differently, particles with small metal cores are better candidates to enhance RE emission. 
This leads to a strong excitation rate increase that compensates the decrease of the apparent quantum yield. 

\begin{table}
\begin{tabular}{cccccccc} 
&d &exc. rate &$\gamma_{rad}/\gamma_0$&$\gamma_{NR}/\gamma_0$ & $\eta$ & fluo. enh. & AEF \\ 
\hline
Rh6G-$Ag(80~\mathrm{nm})@SiO_2$ &4~nm&35& 9.7 &5 &0.66 & 19& 3.8\\
\hline
$\mathrm{Eu^{3+}}$-$Ag(30~\mathrm{nm})@SiO_2$&6~nm&210&1.8 &3 &0.37 & 69 &11.0 \\
\hline

$\mathrm{Eu^{3+}}$-$Ag(100~\mathrm{nm})@SiO_2$&5.5~nm&90&8. &4.2 &0.66 &47 &6.7 
\end{tabular}
\caption{Comparison of excitation and emission rates modification calculated at the optimum distance for Rh6G and $\mathrm{Eu^{3+}}$ 
doped silver core-shell. 
The local fluorescence enhancement is slightly below the product of excitation rate $\times$ apparent quantum yield since it obeys to 
Eq.\ref{eq:alphaMie}.
}
\label{table:Rates}
\end{table}

By increasing metal nanoparticle size, we observe a second optimum size (100 nm) for which efficient exaltation 
of the europium fluorescent signal occurs ($\approx 50$) (Fig.~\ref{fig:AgEu}b). 
In this later case, the excitation field couples to the plasmon quadrupolar mode ($\lambda=415~\mathrm{nm}$) and radiative rate is also 
efficiently enhanced by coupling to the dipolar mode ($\lambda \approx 525~\mathrm{nm}$, Fig.~\ref{fig:AgEu}c). 
This leads to intermediate luminescence 
enhancement as compared to dye-doped system and dipolar assisted RE luminescence enhancement (see table \ref{table:Rates}).
{In addition, this possibility to enhance the excitation by coupling to the quadrupolar quadrupolar and emission by coupling to the dipolar resonance offers a supplementary degree of freedom for luminescence control.  
}

\begin{figure}[h!]
\begin{center}
\includegraphics[width=13cm]{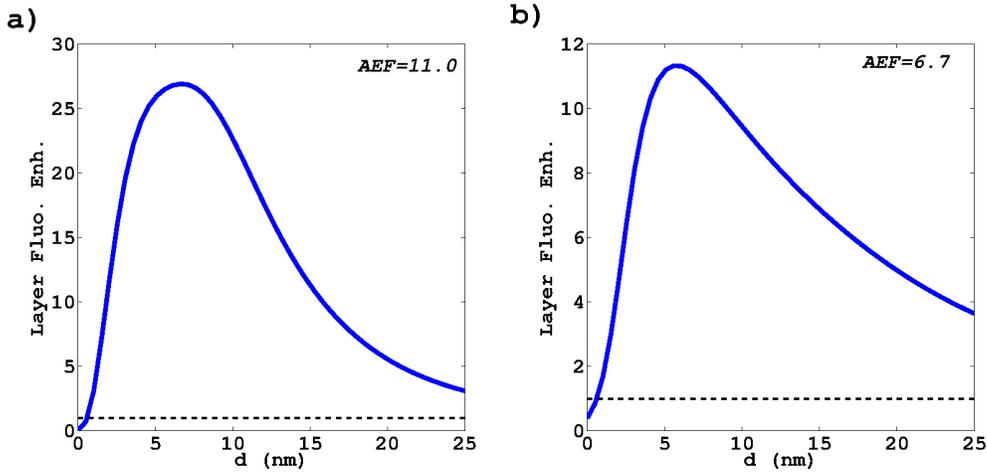}
\caption{\label{fig:AEFAgEu} Layer fluorescence enhancement of a $\mathrm{Eu^{3+}}$ doped core--shell $\mathrm{Ag@SiO_2}$ spherical particle. 
with a silver core of 30 nm (a) or 100 nm (b). The average fluorescence enhancement of the whole doped layer (25 nm) 
is reported on each figure. The horizontal line indicates the enhancement/inhibition threshold.}
\end{center}
\end{figure}  
Having determined the optimal nanoparticle sizes in order to enhance red luminescent of a single $\mathrm{Eu^{3+}}$ ion, 
we now estimate the enhancement of an infinitely thin doped shell. 
Figure \ref{fig:AEFAgEu} shows the evolution of the layer fluorescence enhancement ${\alpha}_{layer}$ 
with distance between metal and emitters and the average fluorescence enhancement factor 
for the whole $\mathrm{Eu^{3+}}$ doped shell for these two optimal sizes. 
For a metal core 30 nm and doped shell thickness of 25 nm, we achieve a strong $AEF=11$. 
This high AEF relies on the strong field enhancement at the plasmon dipolar mode resonance. 
Quenching is limited to emitters very close to the metal surface. 

{Finally, it is not possible to achieve fluorescence enhancement for these rare earth doped systems using a gold core 
(even optimizing the core diameter) since the gold particle dipolar resonance cannot be tuned to the $\mathrm{Eu^{3+}}$ absorption peak $\lambda=415~\mathrm{nm}$ 
(see Fig.~\ref{fig:AuAgFluoDisp}).  
}
\subsection{Emission in the near infrared}
\subsubsection{Spherical core--shell particle}
\begin{figure}[h!]
\begin{center}
\includegraphics[width=13cm]{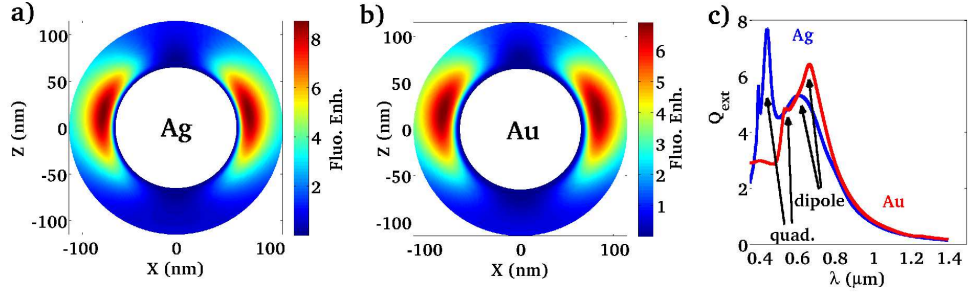}
\caption{\label{fig:ErAuorAgX800Em1550}Fluorescence enhancement of randomly oriented $\mathrm{Er^{3+}}$ near a 130 nm silver (a) 
or gold (b) particle. The excitation and emission wavelength are 
$\lambda_{exc}=800~ \mathrm{nm}$ and $\lambda_{em}=1.55~\mu m$, respectively. 
c) Extinction efficiency of a 130 nm silver (blue line) or gold (red line) particle. 
The embedding medium is $\mathrm{SiO_2}$ (optical index $n=1.5$).  }
\end{center}
\end{figure}  
Erbium ions ($\mathrm{Er^{3+}}$) are widely used for optical amplification for telecom applications since they 
emit around $\lambda_{em}=1.55~\mu m$. We investigate the possibility to enhance 
their IR emission signal using core-shell configuration. In Erbium Doped Fiber Amplifier (EDFA), 
they are usually pumped at 800 or 980 nm wavelength 
in order to limit non radiative losses present with high energy pumping. 
We consider excitation wavelength $\lambda_{exc}=800~\mathrm{nm}$ and emission at $\lambda_{em}=1.55~\mu m$. 
We calculate a maximum enhancement of spherical core-shell doped with $\mathrm{Er^{3+}}$ for a 130 nm metal core, 
either for $\mathrm{Ag@SiO_2}$ or $\mathrm{Au@SiO_2}$. We find a maximum local enhancement up to 8.5(6.5) fold near a silver (gold) particle 
resulting from the excitation of the large dipolar resonance 
(see Fig.~\ref{fig:ErAuorAgX800Em1550}c). 
Our simulation gives then an average enhancement factor of 2.3 (1.7) for the whole doped volume $V_0$ 
for a silver (gold) core coated with a 25 nm doped shell layer. 
This poor AEF is due to the position of excitation and emission wavelength of $\mathrm{Er^{3+}}$, both far from plasmon maximum. 
Since it is possible to red-shift the longitudinal dipolar plasmon resonance with nanoparticle presenting high aspect ratio, 
we propose to optimize the AEF with an elongated metal core.

\subsubsection{Nanorod core--shell particle}
Spherical core--shell particles present limited AEF and imposes to use large particle in order to red shift the plasmon dipolar resonance 
to match the $\mathrm{Er^{3+}}$ absorption spectrum. Another possibility is to use rod shape particle with high aspect ratio 
\cite{Mertens-Polman:2006,Mertens-Polman:2009,Liaw-Tsai:2012}. In this last section, we investigate  core-shell nanorod and estimate 
the fluorescent enhancement [Eqs~(\ref{eq:alpha}) and (\ref{eq:AEF})]. 
In this goal, the excitation electric field ${\bf  E}$ and emission rates $\Gamma_{i}$ need to be evaluated at any location 
in the doped layer. In case of arbitrary structure, they can be calculated thanks to the 
Green's dyad technique \cite{Girard:2005,PRBBaffou:2008,Girard-Dujardin-Baffou-Quidant:2008}. 
Liaw {\it et al} investigate similar structures using multiple multipole method \cite{Liaw-Tsai:2012}. 
However, they limit the computation to a plane (emitter orientation is constrained in 2 dimensions and only the doped layer 
inside the plane of incidence is considered). Since the highest fluorescence rate is obtained for these emitters positions 
and orientations, this leads to overestimate the enhancement factor of the whole doped layer. 
There work however paves the way to optimize the shape of the elongated core-shell particle. 
% Let us also mention the experimental work of 
% Mertens and Polman who measured a two fold luminescence enhancement for $\mathrm{Er^{3+}}$ near silver nanorods array 
% (100 nm $\times$ 400 nm rods). The nanorods were designed so that their dipolar resonance match the emission wavelength of 
% the $\mathrm{Er^{3+}}$ ($\lambda_{em}=1.54~\mathrm{\mu m}$) and not the pump wavelength ($\lambda_{exc}=532~\mathrm{nm}$).

\begin{figure}[h!]
\begin{center}
\includegraphics[width=12cm]{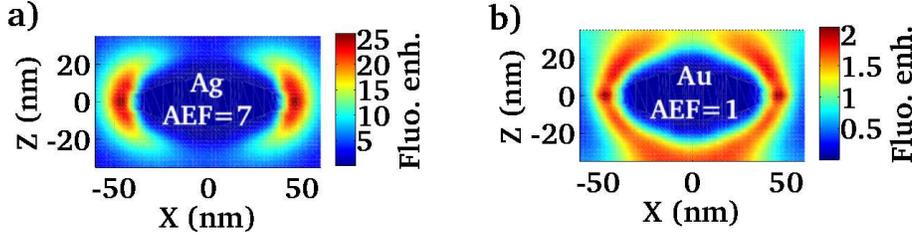}
\caption{\label{fig:AuRodEr}Fluorescence enhancement of randomly oriented $\mathrm{Er^{3+}}$ near a silver (a) or gold (b) nanorod ($70~\mathrm{nm}~\times~20~\mathrm{nm}$). 
The incident electric field is polarized along the rod long axis. The excitation and emission wavelength are $\lambda_{exc}=800~\mathrm{nm}$ and 
$\lambda_{em}=1.55~\mu m$, respectively. The AEF over the whole (3D) 25 nm doped shell is indicated. The two first nm of the shell are 
undoped ($\mathrm{SiO_2}$ spacer). The embedding medium is $\mathrm{SiO_2}$ (optical index $n=1.5$)}
\end{center}
\end{figure}

Figure \ref{fig:AuRodEr}(a) represents the fluorescent enhancement near a silver nanorod. The aspect ratio of the rod has been fixed 
in order to match the particle resonance and the excitation field ($\lambda_{exc}=800~\mathrm{nm}$). The volume of the nanorod is the same as  
for a 30 nm spherical particle so that the achieved enhancement factor can be compared to the visible regime (Fig. \ref{fig:AgEu}a).  
The tip effect strengthens the field exaltation at the dipolar resonance and we observe a fluorescence 
enhancement up to 25 fold near the rod tip. Finally, we achieve an average enhancement factor $AEF=7$ for the whole doped shell, 
comparable to the visible regime (Fig.~\ref{fig:AEFAgEu}). There is no improvement for a gold core ($AEF=1$, Fig.~\ref{fig:AuRodEr}b) {due to high losses. Nevertheless, we would like to mention that we use the dielectric constant of bulk metal although crystalline nanoparticles present lower losses. This could lead to small improvement of the enhancement factor.}

\section{Conclusion}
In this work, we have quantified the plasmon contribution to the luminescence enhancement of rare earth doped metallic core-shell nanoparticles. 
{
This would therefore help in discriminating and then optimizing the different enhancement mechanisms that could play a role in rare-earth doped plasmonic core-shell particles.}

The highest {plasmon mediated} enhancement is achieved when the particle dipolar resonance matches the excitation wavelength. 
The enhancement mainly comes from excitation rate exaltation by coupling to the dipolar mode whereas emission process is weakly modified. {It is also possible to enhance the excitation by coupling to the quadrupolar plasmon and emission by coupling to the dipolar plasmon. 
This offers a supplementary degree of freedom for luminescence control.} 
These results differ from dye-doped particles. Indeed, the small Stokes shift between emission and excitation wavelengths leads then to choose a particle resonance overlapping the two in order to enhance both excitation and radiative rates by coupling to the dipolar resonance. 

We demonstrate average enhancement factor of the fluorescence on the overall RE-doped silver nanoparticle $AEF=11$ and $AEF=7$ 
in the visible and near-infrared regime, respectively. Gold core leads to lower effect due to larger losses.   

These values are similar to the AEF measured on RE-metal nanocluster system \cite{Ma-Dosev-Kennedy:2009,Eichelbaum:2009} 
where the enhancement originates from energy transfer. 
Up to 250 fold enhancement has been reported but probably due to concentration effect \cite{Eichelbaum:2009}. 
We however envision different applications for RE-doped core-shell plasmonic particles and RE-metal nanoclusters. 
Colloidal solutions of plasmonic particles can be synthesized and would benefit as {\it e.g} biolabels or solar cells whereas RE-nanocluster 
doped glasses are promissing materials for telecom fiber amplification. 

\section*{Acknowledgement}
This work is supported by the Agence Nationale de la Recherche (Fenopti$\chi$s ANR-09-NANO-23 and HYNNA ANR-10-BLAN-1016).  
Calculations were performed using DSI-CCUB resources (Universit\'e de Bourgogne).



%% References


%\bibliographystyle{elsarticle-num}
%\bibliography{/home/gcolas/Bureau/ARTICLE/nanolib.bib,/home/gcolas/Bureau/ARTICLE/ref.bib,/home/gcolas/Bureau/ARTICLE/ldos.bib,/home/gcolas/Bureau/ARTICLE/SPPguide.bib,/home/gcolas/Bureau/ARTICLE/mol.bib,/home/gcolas/Bureau/ARTICLE/biblio.bib,/home/gcolas/Bureau/ADMINISTRATIF/CV/CVtex0/DONNEES/refGCF.bib}

\begin{thebibliography}{10}
\expandafter\ifx\csname url\endcsname\relax
  \def\url#1{\texttt{#1}}\fi
\expandafter\ifx\csname urlprefix\endcsname\relax\def\urlprefix{URL }\fi
\expandafter\ifx\csname href\endcsname\relax
  \def\href#1#2{#2} \def\path#1{#1}\fi



\bibitem{Timmerman:2008}
D. Timmerman, I. Izeddin, P. Stallinga, I. Yassievich, T. Gregorkiewicz,  
Space-separated quantum cutting with silicon nanocrystals for photovoltaic applications, Nature Photonics 2 (2008) 105.

\bibitem{Bouzigues:2011}
C. Bouzigues,  T. Gacoin, A. Alexandrou,
Biological Applications of Rare-Earth Based Nanoparticles,
ACS Nano 11 (2011) 8488.

\bibitem{Wang:2011}
C. Wang,H. Tao, L. Cheng, Z. Liu
Near-infrared light induced in vivo photodynamic therapy of cancer based on upconversion nanoparticles,
Biomaterials 32 (2011) 6145.

\bibitem{Bunzli:2007}
J.C. Bunzli, S. Comby, A.S. Chauvin, C. Vandevyver, 
New Opportunities for Lanthanide Luminescence, J. Rare Earths 25  (2007) 257.


\bibitem{Pettinger:2010}
B.~Pettinger, Single-molecule surface- and tip-enhanced Raman spectroscopy, 
Molecular Physics 108 (2010) 2039--2059.

\bibitem{Fort-Gresillon:2008}
E. ~Fort and S.~Gr\'esillon, 
Surface enhanced fluorescence, J. Phys. D: Appl. Phys. 41 (2008), 013001.


\bibitem{Giannini-FrenadezDominguez-Heck-Maier:2011}
V.~Giannini, A.~I. Fernandez-Dominguez, S.~C. Heck, S.~A. Maier, Plasmonic
  nanoantennas: Fundamentals and their use in controlling the radiative
  properties of nanoemitters, Chemical Reviews 111 (2011) 3888--3912.

\bibitem{Sun-Khurgin-Tsai:2012}
G.~Sun, J.~B. Khurgin, D.~P. Tsai, Comparative analysis of photoluminescence
  and raman enhancement by metal nanoparticles, Optics Letters 37 (2012)
  1583--1585.

\bibitem{Fang-Seong-Dlott:2008}
Y.~Fang, D.~D. Seong, N.-H.;~Dlott, Measurement of the distribution of site
  enhancements in surface-enhanced raman scattering, Science 321 (2008) 388.

\bibitem{bharadwaj07b}
P.~Bharadwaj, L.~Novotny, Spectral dependence of single molecule fluorescence
  enhancement, Opt. Expr. 15 (2007) 14266--14274.

\bibitem{Strohhofer-Polman:2002}
C. Strohh{\"o}fer, A. Polman, Silver as sensitizer for erbium, 
Applied Physics Letters 81 (2002) 1414. 

\bibitem{Mertens-Polman:2006}
H. Mertens, A. Polman, Plasmon-enhanced erbium luminescence, 
Applied Physics Letters 89 (2006) 211107. 

\bibitem{Marques-Almeida:2007}
A.C.Marques, R.M. Almeida, 
Er photoluminescence enhancement in Ag-doped sol-gel planar waveguides, 
Journal of Non-Crystalline Solids 353 (2007) 2613-2618.

\bibitem{Aslan-Wu-Lakowics-Geddes:2007}
K.~Aslan, M.~Wu, J.~R. Lakowicz, C.~D. Geddes, Fluorescent core-shell $\mathrm{Ag@SiO_2}$
  nanocomposites for metal-enhanced fluorescence and single nanoparticle
  sensing platforms, J. Am. Chem. Soc. 129 (2007) 1524--1525.

\bibitem{Ma-Dosev-Kennedy:2009}
 Z. Ma, D. Dosev and I. M. Kennedy, A microemulsion preparation of
nanoparticles of europium in silica with luminescence enhancement using silver
Nanotechnology 20 (2009) 085608 
 
\bibitem{Kassab-Araujo:2010}
L. R. P. Kassab, D. S. da Silva, C. B. de Ara\'ujo, 
Influence of metallic nanoparticles on electric-dipole and magnetic-dipole transitions of $\mathrm{\mathrm{Eu^{3+}}}$ doped germanate glasses, 
Journal of Applied Physics 107 (2010) 113506.

\bibitem{Som-Karmakar:2011}
T.~Som, B.~Karmakar, Nano silver:antinomy glass hybrid nanocomposites and their enhanced fluorescence application
Solid State Sciences 13 (2011) 887--895.

\bibitem{vanWijngaarden-Meijerink:2011}
J.~T. van Wijngaarden, M.~M. van Schooneveld, C.~de~Mello~Donega, A.~Meijerink,
  Enhancement of the decay rate by plasmon coupling for $\mathrm{\mathrm{Eu^{3+}}}$ in an Au
  nanoparticle model system, EPL 93 (2011) 57005.

\bibitem{Reisfeld-Gaft:2011}
R.~Reisfeld, T.~Saraidarov, G.~Panzer, V. ~Levchenko, M. Gaft, New optical material europium
EDTA complex in polyvinyl pyrrolidone films with fluorescence enhanced by
silver plasmons, Optical Materials 34 (2011) 351--354. 

\bibitem{Deng-Goldys:2012}
W. Deng, L. Sudheendra, J. Zhao, J. Fu, D. Jin, I.M. Kennedy, E.M Goldys, 
Upconversion in NaYF4:Yb, Er nanoparticles amplified by metal nanostructures, 
Nanotechnology 22 (2011) 325604.


\bibitem{Rivera-Nunes-Marega Jr:2012}
V.A.G.~Rivera, Y.~Ledemi, S.P.A.~Osorio, D.~Manzani, Y.~Messaddeq, L.A.O.~Nunes, E.~Marega Jr., 
Efficient plasmonic coupling between $\mathrm{\mathrm{Er^{3+}}}$:(Ag/Au) in tellurite glasses, 
Journal of non-crystalline solids 358 (2012) 399--405.

\bibitem{Amjad:2013}
R.~Amjad, M.~Sahar, M.~Dousti, S.~Ghoshal and M.~Jamaludin, 
Surface enhanced Raman scattering and plasmon enhanced fluorescence in zinc-tellurite glass, 
Optics Express 21 (2013), 21282--21290.


\bibitem{Malta-CoutoDosSantos:1990}
O.L.~Malta, M.A.~Couto dos Santos, 
Theoretical analysis of the fluorescence yield of rare earth ions in glasses containing small metallic particles, 
Chemical Physics Letters 174 (1990) 13-18.

\bibitem{Greffet:2009}
R. Esteban,1, M. Laroche and J.-J. Greffet
Influence of metallic nanoparticles on upconversion processes
Journal of Applied Physics 105 (2009), 033107

\bibitem{Fischer-Goldschlmidt:2012}
S. Fischer, F. Hallermann, T. Eichelkraut, G. {von Plessen}, K.W. Kr\"amer, D. Biner,H. Steinkemper, M. Hermle, J.C. Goldschmidt, 
Plasmon enhanced upconversion luminescence near gold nanoparticles-simulation and analysis of the interactions, 
Optics Express (2012) 271-282.


\bibitem{Karaveli:2011}
S. Karaveli, R. Zia, Spectral Tuning by Selective Enhancement  of Electric and Magnetic Dipole Emission
Physical Review Letters 106 (2011) 193004


\bibitem{Eichelbaum:2009}
M.~Eichelbaum, K.~Rademann, Plasmonic enhancement or energy transfer? on the
  luminescence of gold-, silver-, and lanthanide-doped silicate glasses and its
  potential for light-emitting devices, Advanced Functional Materials 19 (2009)
  2045--2052.

\bibitem{Busson-Bidault:2012}
M.~Busson, B.~Rolly, B.~Stout, J.~W. N.~Bonod, S.~Bidault, Photonic engineering
  of hybrid metal-organic chromophores, Angew. Chem. Int. Ed. 51 (2012)
  11083--11087.

\bibitem{Maurizio:2011}
C. Maurizio, E. Trave, G. Perotto, V. Bello, D. Pasqualini, P. Mazzoldi, G. Battaglin, T. Cesca, C. Scian, G. Mattei, 
Enhancement of the $\mathrm{\mathrm{Er^{3+}}}$ luminescence in Er-doped silica by few-atom metal aggregates, 
Physical Review B 83 (2011) 195430.

\bibitem{Xiong:2012}
B. Peng, Q. Zhang, X. Liu, Y. Ji, H. Demir, C. Huan, T. Sum and Q. Xiong,
Fluorophore-Doped Core Multishell Spherical Plasmonic Nanocavities: Resonant Energy Transfer toward a
Loss Compensation, ACS Nano 6 (2012), 6250--6259.


\bibitem{Liaw-Kuo:2010}
J.-W. Liaw, C.-L. Liu, W.-M. Tu, C.-S. Sun, M.-K. Kuo, Average enhancement
  factor of molecules-doped coreshell ($\mathrm{Ag@SiO_2}$) on fluorescence, Optics Express
  18 (2010) 12788--12797.

\bibitem{Johnson-Christy:1972}
P.~Johnson, R.~Christy, Optical constants of the noble metals, Physical Review
  B 6 (1972) 4370--4379.

\bibitem{OptExpGCF:2008}
G.~{Colas des Francs}, A.~Bouhelier, E.~Finot, J.-C. Weeber,
  A.~Dereux, C.Girard, E.~Dujardin, Fluorescence relaxation in the near-field
  of a mesoscopic metallic particle: distance dependence and role of plasmon
  modes, Optics Express 16 (2008) 17654-- 17666.

\bibitem{JCPGCF:2002}
{G. {Colas des Francs}}, C.~Girard, A.~Dereux, Theory of near-field
  optical imaging with a single molecule as a light source, Journal of Chemical
  Physics 117 (2002) 4659--4666.

\bibitem{PRELevequeGCF:2002}
G. L\'ev\^eque \emph{et~al.}
Polarization state of the optical near-field, Physical Review E 65 (2002), 36701.

\bibitem{Kim-Leung-George:1988}
Y.~S. Kim, P.~T. Leung, T.~F. George, Classical decay rates for molecules in
  the presence of a spherical surface: a complete treatment, Surface Science
  195 (1988) 1--14.

\bibitem{Sinzig-Quinten:1995}
J.~Sinzig, M.~Quinten, Scattering and absorption by spherical multilayer
  particles, Applied Physics A 58 (1994) 157--162.

\bibitem{Reineck:2013}
V.~Reineck, D.~G\'omez,S.~Ng, M.~Karg,T.~Bell, P.~Mulvaney and U.~Bach, 
Distance and Wavelength Dependent Quenching of Molecular Fluorescence by 
$\mathrm{Au@SiO_2}$ Core-Shell Nanoparticles, 
ACS Nano 7 (2013) 6636.

\bibitem{Hartling-Reichenbach-Eng:2007}
T.~H{\"a}rtling, P.~Reichenbach, L.~M. Eng, Near-field coupling of a single
  fluorescent molecule and a spherical gold nanoparticle, Optics Express 15
  (2007) 12806--12817.

\bibitem{Pillonnet-Jurdyc:2012}
A.~Pillonnet, A.~Berthelot, A.~Pereira, O.~Benamara, S.~Derom,
  G.~{Colas des Francs}, A.-M. Jurdyc, Coupling distance between $\mathrm{\mathrm{Eu^{3+}}}$
  emitters and Ag nanoparticles, Applied Physics Letters 100 (2012) 153115.

\bibitem{Mertens-Polman:2009}
H.~Mertens, A.~Polman, Strong luminescence quantum-efficiency enhancement near
  prolate metal nanoparticles: Dipolar versus higher-order modes, Journal of
  applied physics 105 (2009) 44302.

\bibitem{Liaw-Tsai:2012}
J.-W. Liaw, H.-Y. Tsai, Theoretical investigation of plasmonic enhancement of
  silica-coated gold nanorod on molecular fluorescence, Journal of Quantitative
  Spectroscopy an Radiative Transfer 113 (2012) 470--479.

\bibitem{Girard:2005}
C.~Girard, Near-field in nanostructures, Report on Progress in Physics 68
  (2005) 1883--1933.

\bibitem{PRBBaffou:2008}
G.~Baffou, C.~Girard, E.~Dujardin, {G. {Colas des Francs}},
  O.~Martin, Molecular quenching and relaxation in a plasmonic tunable nanogap,
  Physical Review B 77 (2008) 121101(R).

\bibitem{Girard-Dujardin-Baffou-Quidant:2008}
C.~Girard, E.~Dujardin, G.~Baffou, R.~Quidant, Shaping and manipulation of
  light fields with bottom-up plasmonic structures, New Journal of Physics 10
  (2008) 105016.

\end{thebibliography}





% \end{thebibliography}

\end{document}